\documentclass[aip,jcp,amsmath,amssymb,reprint]{revtex4-1}
\usepackage{graphicx}
\usepackage{amssymb}
\usepackage{amsmath}
\usepackage{color}
\usepackage[]{hyperref}
\usepackage[caption=false,justification=raggedright]{subfig}
\usepackage{bm}
\usepackage{upgreek}
\usepackage[multidot]{grffile}
\usepackage{float}
\usepackage{relsize}
\usepackage{rotating}
\usepackage{array}
\usepackage{tikz}
\usepackage{tabularx}
\usepackage[utf8]{inputenc}
\usepackage{upgreek}
\usetikzlibrary{backgrounds}
\usepackage{standalone}
\usepackage{epstopdf}
\usetikzlibrary{calc, decorations,decorations.text,decorations.pathmorphing,shapes.callouts,shapes.symbols}

\tikzset{fontscale/.style = {font=\relsize{#1}}}

\begin{document}

\title{Dipolar Capillary Interactions between Tilted Ellipsoidal Particles Adsorbed at Fluid-Fluid Interfaces}

\author{Gary B. Davies}
\email{gbd@icp.uni-stuttgart.de}
\affiliation{Institute for Computational Physics, Allmandring 3, 70569 Stuttgart, Germany.}

\author{Lorenzo Botto}
\email{l.botto@qmul.ac.uk}
\affiliation{School of Engineering and Materials Science, Queen Mary, University of London, London E1 4NS, United Kingdom.}

\newcommand{\degr}{^{\circ}}
\newcommand{\mba}{bond angle maximum }

\begin{abstract}
Capillary interactions have emerged as a tool for the directed assembly of particles adsorbed at fluid-fluid interfaces, and play a role in controlling the mechanical properties of emulsions and foams. In this paper, following Davies \emph{et al}.~[\emph{Advanced Materials}, 26, 6715 (2014)] investigation into the assembly of ellipsoidal particles at interfaces interacting via dipolar capillary interactions, we numerically investigate the interaction between tilted ellipsoidal particles adsorbed at a fluid-fluid interface as their aspect ratio, tilt angle, bond angle, and separation vary. High-resolution Surface Evolver simulations of ellipsoidal particle pairs in contact reveal an energy barrier between a metastable tip--tip configuration and a stable side--side configuration. The side--side configuration is the global energy minimum for all parameters we investigated. Lattice Boltzmann simulations of clusters of up to 12 ellipsoidal particles show novel highly symmetric flower-like and ring-like arrangements. 
\end{abstract}


\maketitle

The study of capillary interactions between particles adsorbed at fluid-fluid interfaces has attracted significant attention in recent years. Capillary interactions play a role in, for example, colloidal assembly,~\cite{loudet_capillary_2005, lumay_self-assembled_2013, ni_insights_2015, isa_particle_2010} interface rheology,~\cite{madivala_exploiting_2009, basavaraj_packing_2006, madivala_self-assembly_2009, kralchevsky_particles_2001} and emulsion/foam stability.~\cite{kralchevsky_capillary_2000,  daware_emulsions_2015, koos_capillary_2014}

Capillary interactions are caused by overlapping particle-induced interface deformations. They can be analysed in terms of different deformation modes, each corresponding to separate terms in the multipole expansion of the particle-induced capillary disturbance. \cite{danov_interactions_2005,botto_capillary_2012} A particle whose weight is comparable to surface tension forces (i.e. has a finite Bond number) induces a capillary monopole.

The interaction between capillary monopoles has been studied extensively.~\cite{nicolson_interaction_1949, chan_interaction_1981, vella_cheerios_2005, danov_electrodipping_2004} Capillary monopoles can be easily experimented with by spreading millimetric particles on planar fluid interfaces. The particles create downward distortions proportional to the particle weight, driving a phenomenon of capillary aggregation that is often referred to as the ``Cheerios effect''.~\cite{vella_cheerios_2005} 

Quadrupolar capillary interactions can arise due to particle geometry: for anisotropic particles, the interface must deform in order to satisfy Young's uniform contact angle boundary condition.~\cite{loudet_capillary_2005} Surface roughness and surface chemical heterogeneity also cause spherical particles to deform the interface in a quadrupolar manner.~\cite{lucassen_capillary_1992, stamou_long-range_2000, xie_tunable_2015, kumar_amphiphilic_2013, park_janus_2011}

A characteristic of both monopolar and quadrupolar capillary interactions is that the strength of their interactions cannot be dynamically tuned easily. For monopolar interactions, the capillary force depends on the particle weight, size, and surface-tension. For quadrupolar interactions, the magnitude of the capillary force is proportional to the surface tension and the size of the particle via a pre-factor that depends only on the contact angle and the particle geometry. None of these properties can be easily or precisely controlled during the course of an experiment. 
\\\indent Recent work using spherical magnetic particles has significantly improved the ability to control the assembly of spherical particles at interfaces by tuning the interplay between magnetic repulsion and monopolar capillary attractions.~\cite{lumay_self-assembled_2013, vandewalle_mesoscale_2013} However, until recently,~\cite{davies_interface_2014, davies_assembling_2014} control of the assembly of ellipsoidal particles, which enable enable the possibility of directed assembly due to their anisotropy, has been lacking.
\\\indent Davies \emph{et al}.~\cite{davies_interface_2014, davies_assembling_2014} recently showed that, by applying a magnetic field perpendicular to a fluid-fluid interface covered with ellipsoidal particles with magnetic moments aligned along their major-axes, the particles tilt and induce dipolar capillary interface deformations.~\cite{davies_interface_2014, davies_assembling_2014} Their simulations revealed that particles align with their nearest neighbours side--side into chains, but chains face other chains such that their particles arrange tip--tip. They also observed straighter, more rigid chains as the particle tilt angles increased. Finally, they showed how to switch off these dipolar capillary interactions by exploiting a previously discovered first-order orientation phase transition in which the particles flip from a tilted to a vertical orientation at a critical dipole-field strength, $B_c$, and corresponding critical tilt angle, $\psi_c$.~\cite{bresme_orientational_2007, bresme_computer_2008, davies_interface_2014, davies_assembling_2014} They did not investigate the energy landscape for dipolar capillary interactions in detail.
\\\indent Newton {\it et al}.~\cite{newton_influence_2014} recently carried out Surface Evolver (SE) simulations of tilted ellipsoidal particles adsorbed at interfaces. Their highly accurate SE simulations characterised the first-order phase transition much more accurately than previous studies.~\cite{davies_interface_2014, bresme_orientational_2007} They found that the critical tilt angle decreases and the critical dipole-field strength increases as the particle aspect ratio increases. They also found that the critical dipole-field strength and critical tilt angle decrease as the particle becomes less neutrally wetting. From an applications perspective, perhaps their most interesting discovery was the significant hysteresis that tilted ellipsoidal particles exhibit due to the nature of the first-order phase transition. 
 
In this paper, we numerically investigate the interaction between two tilted ellipsoidal particles using Surface Evolver.~\cite{brakke_surface_1992} The tilting causes interface deformations and hence capillary interactions between the particles. We find that the features of these capillary interactions, which are dipolar in nature, are quite different from the more commonly studied case of monopolar and quadrupolar capillary interactions. 

An advantage of SE versus methods based on fixed Eulerian grids such as lattice Boltzmann is its superior accuracy in the evaluation of areas, and therefore surface energies, due to its ability to use fine and non-uniform surface meshes. To investigate many body effects, we use lattice Boltzmann simulations to study the equilibrium configurations for clusters of up to 12 ellipsoidal particles.  The simulation results enable us to discuss some of the local micro structural features present in the monolayers of ellipsoidal particles simulated by~\citet{davies_interface_2014} and the limitations of pair interaction predictions.

We show that the side--side configuration is indeed the lowest energy configuration for two tilted ellipsoidal particles in contact. Additionally, we find that an energy barrier exists between the side--side and tip--tip configurations, and that this energy barrier increases with particle tilt angle. This increase in the free energy barrier explains the increased rigidity of chains with increasing tilt angle observed by~\citet{davies_assembling_2014} Finally, we theoretically develop a far-field pair potential between two tilted ellipsoidal particles, which we validate with numerical simulations. 

\section{Methods}
%%%%%%%%%%%%%%%%%%
\begin{figure}
 \includegraphics[width=\linewidth]{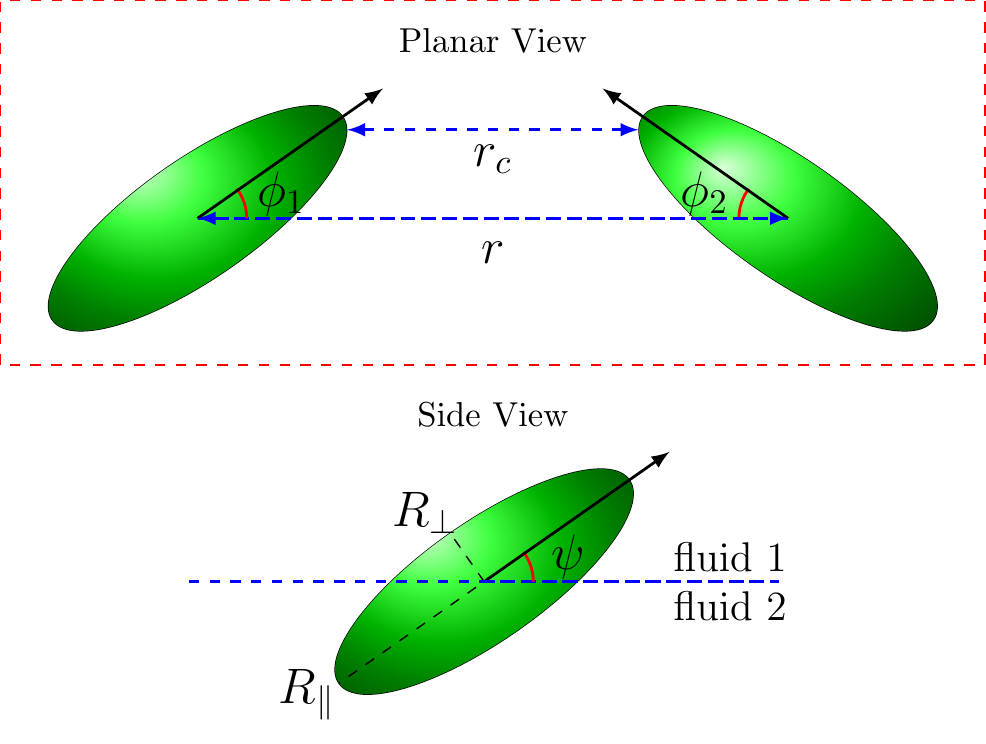}
 \caption{Planar view of both particles and side view of a single particle. The particles have aspect ratio $\alpha=R_{\|}/R_{\perp}$, where $R_{\|}$ and $R_{\perp}$ are the axes parallel and perpendicular to the particle's major axis, respectively. The particles are tilted by an angle $\psi$ with respect to the plane of the undeformed interface, inducing a dipolar capillary deformation, separated by a centre--centre distance $r$ and surface--surface distance $r_c$. In our simulations, each particle has the same tilt angle, and we measure the interaction energy of particles in a mirror symmetric configuration (bond angles $\phi_1=\phi_2=\phi$). }
\label{plot:system_setup}
\end{figure}
%%%%%%%%%%%%%%%%%%

We simulate identical ellipsoidal particles having minor semi-axis $R_{\perp}$, major semi-axis  $R_{\|}$, and aspect ratio  $\alpha=R_{\|}/R_{\perp}$. The contact angle is uniform and equal to $90\degr$. In the first part of the paper pairs of ellipsoidal particles are simulated in mirror-symmetric configurations as a function of the tilt angle $\psi$, the bond angle $\phi$, and the inter-particle separation $r$ (Fig.~\ref{plot:system_setup}). The corresponding minimal surface-surface separation is $r_c$. 

To calculate the interaction energy, we measure the total interface area $A_{12}$ corresponding to a given particle configuration.  The surface free energy for a particle adsorbed at a fluid-fluid interface is ~\cite{davies_detachment_2014, davies_interface_2014,bresme_orientational_2007,bresme_computer_2008,faraudo_stability_2003, aveyard_particle_1996, binks_colloidal_2006}

\begin{align}
E = \gamma_{12} A_{12} + \gamma_{1p}A_{1p} + \gamma_{2p}A_{2p}
\label{eqn:energy}
\end{align}

\noindent where $\gamma_{ij}$ and $A_{ij}$ are the surface energies and contact areas between the $ij^{th}$ phases, respectively ($i,j =$ \{1: fluid $1$, 2: fluid $2$, p: particle\}). Young's relation requires that $\gamma_{12}  \cos \theta = \gamma_{1p}-\gamma_{2p}$, where $\theta$ is the contact angle calculated from fluid $2$. Therefore, up to a constant, $E= \gamma_{12} A_{12} + \gamma_{12} \cos \theta A_{1p}$. In this paper we consider the case $\theta=90\degr$, for which the wetting energy contribution is zero and $E = \gamma_{12} A_{12}$.

To obtain the capillary energy when the particles are in contact, we simulate the interface configuration for small values of $r_c$ and extrapolate $E$  to $r_c = 0$ using quadratic extrapolation. \cite{botto_capillary_2012-1} We calculate the centre--centre separation corresponding to $r_c=0$ using a bisection algorithm that minimises the overlap between the ellipsoidal particles. 
\\\indent Using Surface Evolver,~\cite{brakke_surface_1992} we calculate the equilibrium interface shape that simultaneously satisfies the contact angle boundary condition and the condition of mechanical equilibrium of the particles.~\cite{botto_capillary_2012-1,lewandowski_orientation_2010} The Surface Evolver code that evolves the interface  and ensures the quality of the grid is identical to that used in our previous papers in which cylindrical particles and ellipsoidal particles at interfaces were studied for the particular case of zero tilt angle.~\cite{botto_capillary_2012-1,lewandowski_orientation_2010} The triangulation of the interface adopted is more refined near the particle and coarser far from it. Surface energies are computed by using in the last stages of the surface evolution of the highest-order Lagrangian interpolation offered by Surface Evolver (4th order). This enables high accuracy in the evaluation of the areas. 
\\\indent The lattice Boltzmann simulations of clusters in Sec.~\ref{sectionD} are carried out using the same simulation algorithm as described in~\citet{davies_interface_2014, davies_assembling_2014} For these simulations, we simulate particles with aspect ratio $\alpha=2.0$.
\section{Results and Discussion}

\subsection{Pair interaction: dependence on the bond angle for particles in contact}
\label{results}

%%%%%%%%%%%%%%%%%%
\begin{figure}
\centering
 \subfloat[\label{plot:representative_e}]
{
	 \includegraphics[width=\linewidth]{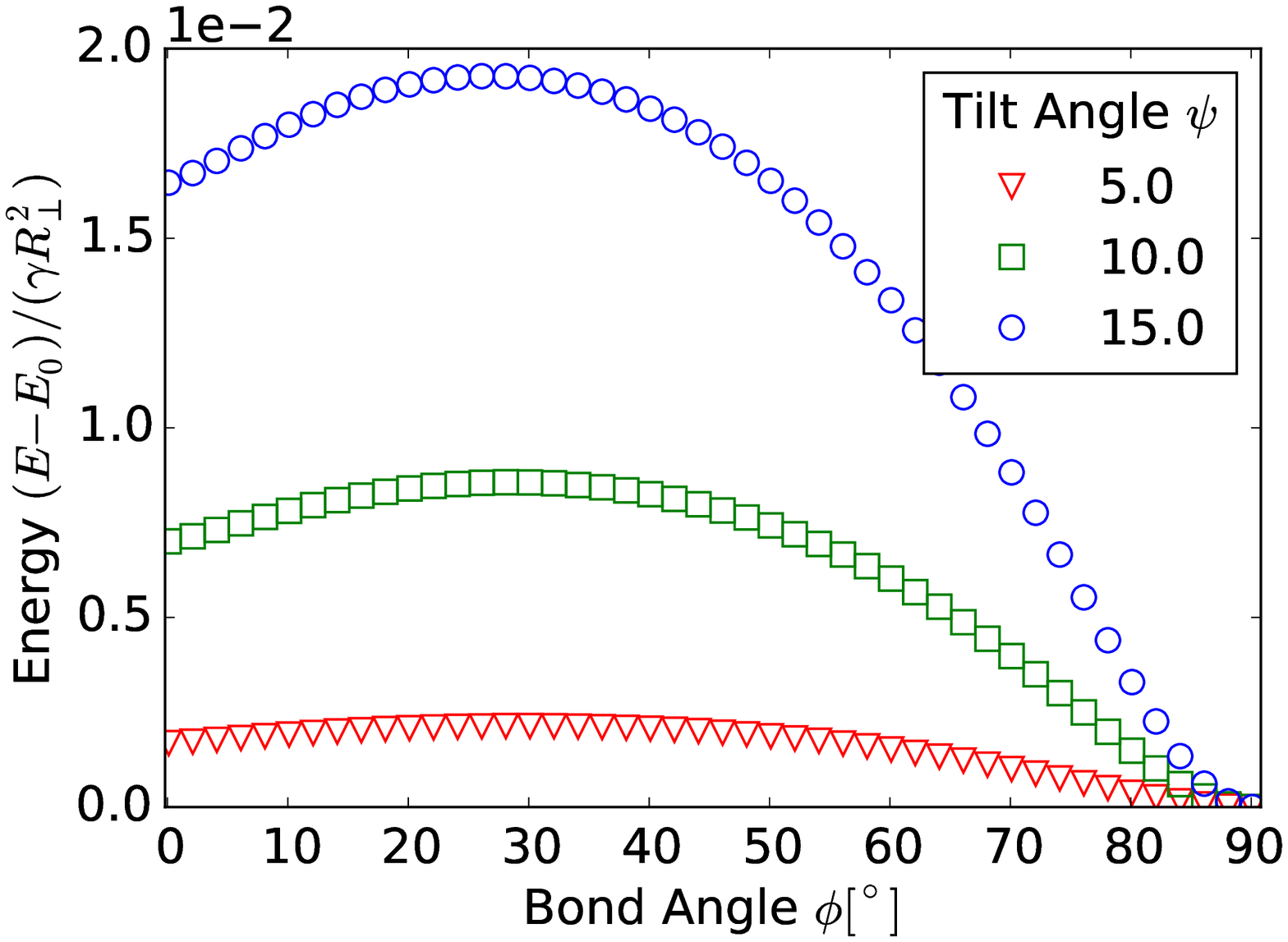} 
	}\\ 
	
	\subfloat[ 
	\label{plot:representative_t}]{
  \includegraphics[width=\linewidth]{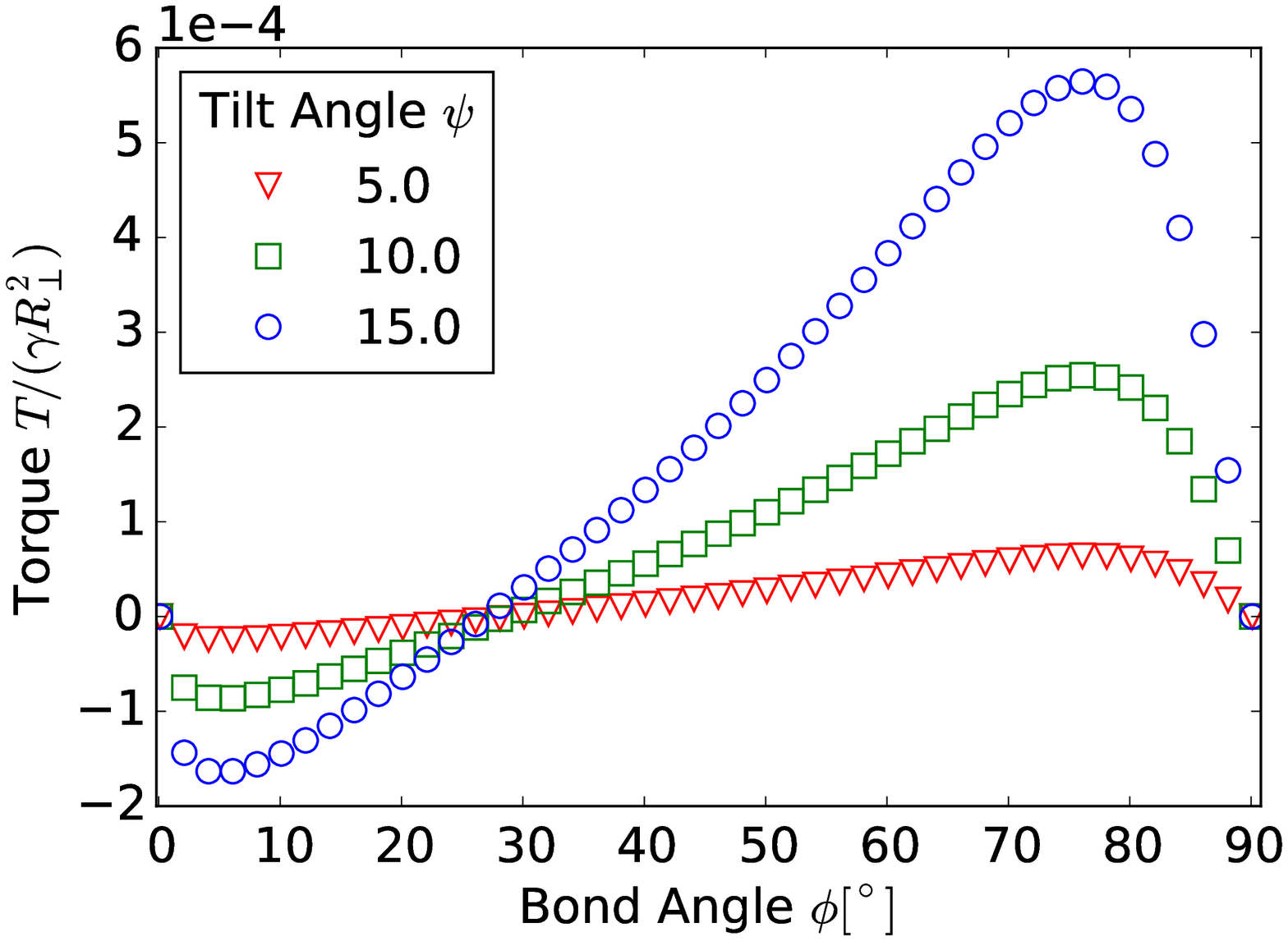}
	}\\ 
\caption{Energy (a) and torque (b) profiles as a function of the bond angle ($\phi = \phi_1 = -\phi_2$) for particles in contact with aspect ratio $\alpha = 2.0$ and tilt angles $\psi = 5\degr$ (red triangles), $10\degr$ (green squares), and $15\degr$ (blue circles). The energy profile in (a) shows that there is an energy barrier between the metastable tip--tip configuration ($\phi=0\degr$) and the global energy minimum side--side configuration ($\phi = 90\degr$). The magnitude of the energy barrier increases with increasing tilt angle.}
	\label{plot:representatives}
\end{figure}
%%%%%%%%%%%%%%%%%%

Fig.~\ref{plot:representative_e} shows the capillary energy profile for two identical ellipsoidal particles with aspect ratio $\alpha = 2.0$ in contact in a mirror symmetric configuration (Fig.~\ref{plot:system_setup}). We plot the capillary energy as a function of the bond angle $\phi = \phi_1 = -\phi_2$ for tilt angles $\psi= 5\degr$, $10\degr$, and $15\degr$. We calculate the energy $E$ with respect to the energy $E_0$ corresponding to the side--side configuration, $\phi = 90\degr$.

\begin{figure}[t]
 \includegraphics[width=\linewidth]{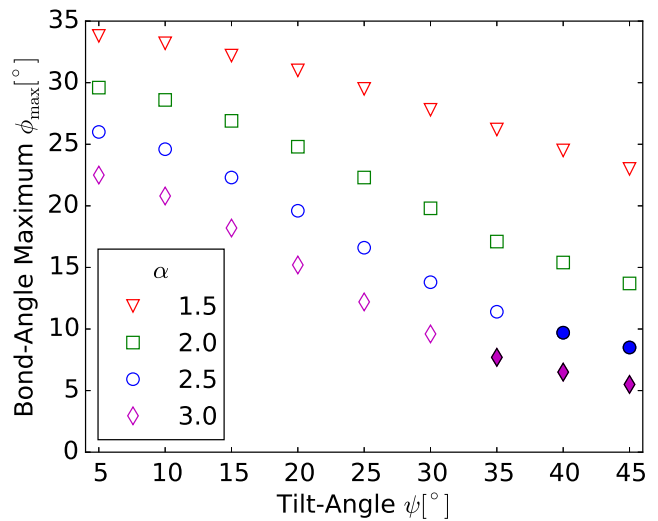}
 \caption{Bond angle maximum $\phi_{max}$ for which the torque is zero as a function of the tilt angle. $\phi_{\mathrm{max}}$ decreases as the tilt angle increases for all aspect ratios $\alpha$. The smaller the aspect ratio, the smaller the change in bond angle maximum as the tilt angle increases, suggesting that particles with smaller aspect ratios have greater possibility to achieve the metastable tip--tip configuration. The filled symbols represent tilt angles beyond the critical tilt angle at which the particle flips into the vertical state.}
\label{plot:Emax_phi_vs_psi}
\end{figure}

Fig.~\ref{plot:representative_e} indicates the presence of a local energy minimum for particles in the tip--tip configuration ($\phi = 0\degr$) and a global energy minimum for particles in the side--side state ($\phi = 90\degr$). The capillary energy is not monotonic: an energy barrier exists that peaks at an angle $\phi_{max}$ and depends on the particle tilt angle for a given aspect ratio. The qualitative features in Fig.~\ref{plot:representatives} are characteristic of all aspect ratios $\alpha$ and tilt angles $\psi$ investigated in this paper. 

The capillary torque $T = - \frac{\partial E}{\partial \phi}$ resisting bond-bending corresponding to Fig.~\ref{plot:representative_e} is shown in Fig.~\ref{plot:representative_t}. As the bond angle increases from $\phi=0\degr$ the torque is negative but increasing until it reaches $T=0$ at $\phi=\phi_{max} $. Therefore, particles with bond angles larger and smaller than $\phi_{max}$ will rotate into the side--side and tip--tip state, respectively.  
 
Fig.~\ref{plot:Emax_phi_vs_psi} shows how $\phi_{max}$ changes with respect to the tilt angle for several different aspect ratios.  For a given tilt angle, increasing the aspect ratio results in a smaller value of $\phi_{max}$ and therefore to a narrower energy well for the tip--tip configuration. This suggests that longer ellipsoidal particles require a smaller angular perturbation to destabilise the tip--tip state. For all aspect ratios, $\phi_{max}$ decreases monotonically as the tilt angle increases. Larger external torques (which can be achieved by increasing the dipole-field strength, $B$, for example) cause larger tilt angles, which will therefore make the side--side configuration even more favourable. 

In addition, Fig.~\ref{plot:Emax_phi_vs_psi} indicates a larger reduction in $\phi_{max}$ for larger aspect ratios as the tilt angle increases. For example: a particle with aspect ratio $\alpha = 3$ has  $\phi_{max} \simeq 22.5\degr$ for  $\psi = 5\degr$ and $\phi_{max}$ decreases to $5\degr$ for a tilt angle $\psi = 45\degr$, a difference of approximately $17.5\degr$. For aspect ratio $\alpha = 1.5$, the equivalent change in the value of $\phi_{max}$ is only $\approx 10\degr$. This indicates that particles with smaller aspect ratios have a larger range of bond angles that lead to a tip--tip configuration than particles with larger aspect ratios. Since the behaviour of tilted ellipsoidal particle monolayers in which the aspect ratio varies has not yet been investigated, this prediction could provide a hint of novel structures in such systems.  

\begin{figure}
 \includegraphics[width=\linewidth]{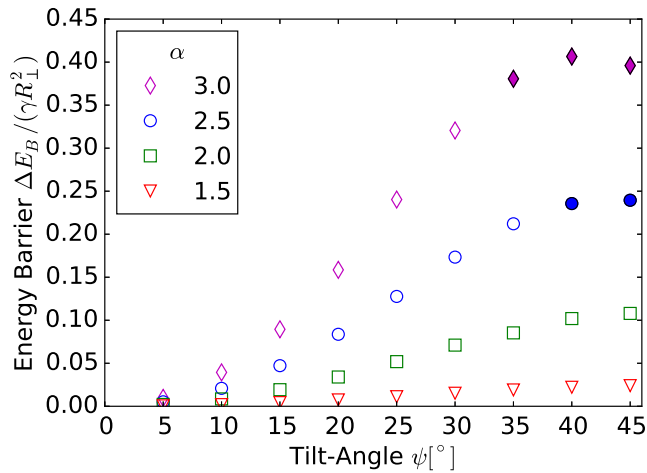}
 \caption{Dependence of the magnitude of the energy barrier $\Delta E_B = E_{\mathrm{max}} - E_{\mathrm{min}}$ on the tilt angle, $\psi$. The energy barrier magnitude increases with tilt angle for all aspect ratios. For a given tilt angle, a larger aspect ratio results in a larger energy barrier. Filled symbols represent tilt angles greater than the critical tilt angle, $\psi_c$, for that aspect ratio. }
\label{plot:energy_barrier}
\end{figure}

In Fig.~\ref{plot:energy_barrier} we characterise the magnitude of the  energy barrier separating the side--side and tip--tip minima by taking the difference between the global energy maximum and minimum, $\Delta E_B = E_{\mathrm{max}} - E_{\mathrm{min}}$. For a given tilt angle, the energy barrier $\Delta E_B$ increases as the aspect ratio increases. It also increases as the tilt angle increases for a given aspect ratio. For aspect ratio $\alpha=3.0$,  $\Delta E_B$  increases until it reaches a maximum at tilt angle $\psi \approx 40\degr$ before decreasing. However, we note that $\psi = 40\degr$ is larger than the critical angle $\psi_c$ at which the particle  transitions into the vertical state when a constant torque is applied.~\cite{davies_interface_2014, newton_influence_2014, bresme_orientational_2007} In our simulations, we fix the tilt angle so that angles larger than $\psi_c$ are therefore permitted; for clarity, in Fig.~\ref{plot:Emax_phi_vs_psi} and Fig.~\ref{plot:energy_barrier} values with filled symbols correspond to $\psi > \psi_c$.

%%%%%%%%%%%%%%%%%%
%%%%%%%%%%%%%%%%%%

For the same particle shape, dipolar and quadrupolar capillary interactions share some qualitative features.  For instance,  in the case of ellipsoidal particles, the side--side configuration is the global energy minimum for both  quadrupolar and dipolar capillary interactions.~\cite{botto_capillary_2012-1} However, for ellipsoidal particles inducing capillary quadrupoles the interaction energy depends monotonically on the bond angle, while an energy barrier is present for ellipsoidal  inducing capillary dipoles, as shown in Fig.~\ref{plot:representatives}. 

\citet{botto_capillary_2012-1} found an energy barrier for cylindrical particles inducing quadrupolar interactions. They attributed this feature to the fact that pairs of cylindrical particles in contact subject to a bond-bending deformation must ``hinge'' at the point of contact, leading to an increased separation between the flat faces of the particles as the bond angle increases. In our case, the energy barrier is rooted in completely different physical features, namely the anisotropic interface distortion induced by tilting. 

\subsection{Pair interaction: dependence on inter-particle separation}

In this section, we present a pair potential between polar capillary dipoles. The derivation is analogous to that used by Stamou {\it et al}.~\cite{stamou_long-range_2000} in their study of quadrupolar interactions, but we replace quadrupoles with dipoles. We invoke the superposition principle, valid in the far-field, which assumes that the interface deformation at any point on the interface is simply the sum of the dipolar interface deformations created by particle $A$ and particle $B$:

\begin{align}
h_{A,B} = H_{A,B} \cos (\phi - \phi_{A, B}) \left(\frac{R_c}{r} \right)
\end{align}

where $H_{A,B}$ are the amplitudes of the dipolar distortions, $R_c$ is the nominal contact line radius, $r$ is the centre--centre separation, and $\phi_{A, B}$ are the particle bond angles. 

Calculating the interface area corresponding to the superposition of two dipoles, and assuming small slopes, gives the following polar dipole pair potential: 

\begin{equation}
E = \gamma_{12} \pi H_A H_B  \frac{R_c^2}{r^2} \cos (\phi_A + \phi_B).
\label{eqn:polardipolepotential}
\end{equation} 

In the system we consider here where the bond angles are symmetric $\phi_A = -\phi_B$ and in the far field the amplitude of the particle-induced interface distortions are identical $H = H_A = H_B$, the interaction energy becomes 

\begin{equation}
E = \gamma_{12} \pi H^2  \frac{R_c^2}{r^2}
\label{eqn:polardipolepotential_simple}
\end{equation} 

\begin{figure}
 \includegraphics[width=\linewidth]{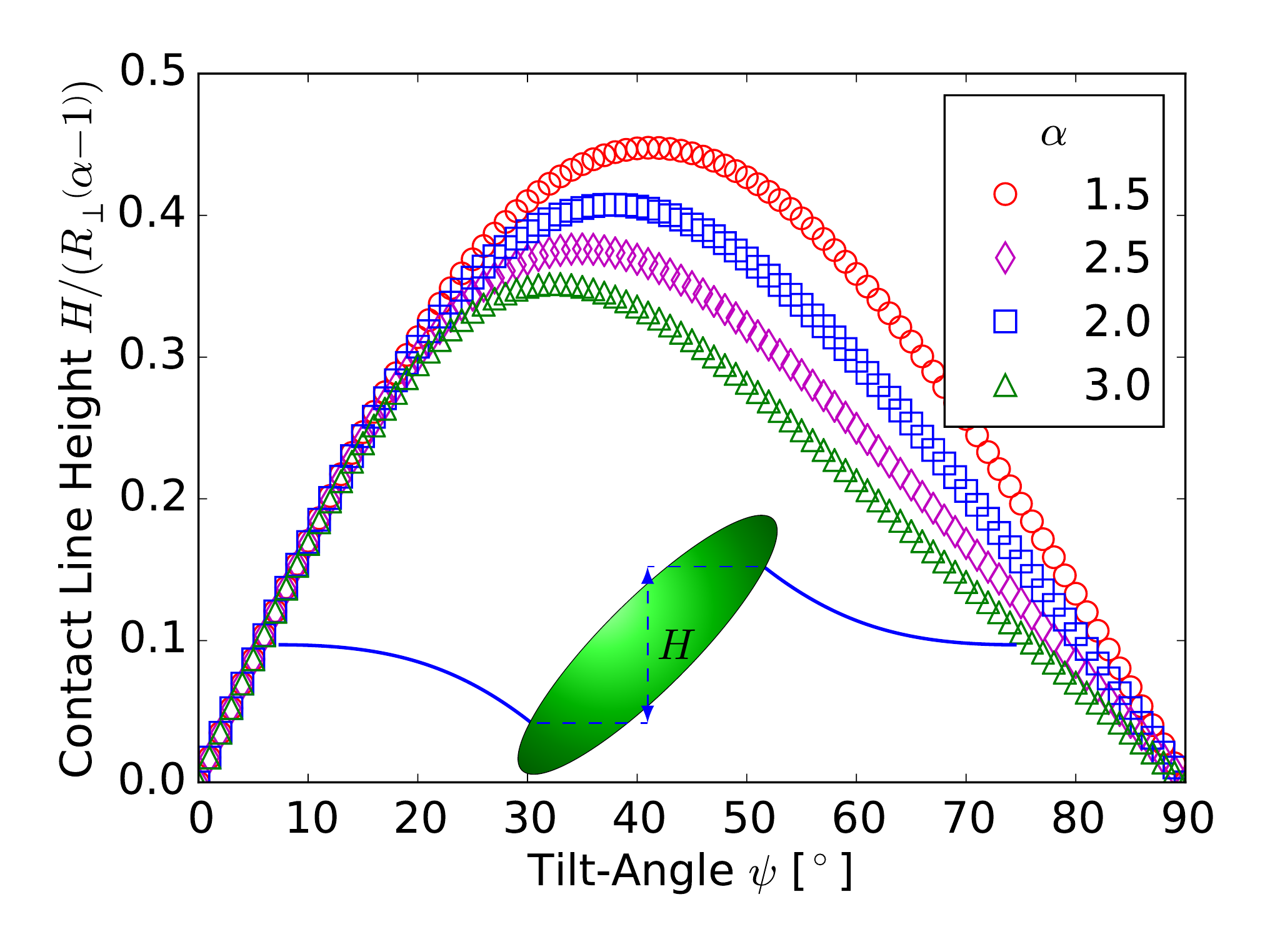}
 \caption{The dependence of the maximal contact line height difference, $H$, on the tilt angle for a single isolated particle with several different aspect ratios, $\alpha$. We normalise the contact line height by $R_{\perp}(\alpha -1)$, finding a scaling law for the small tilt angle regime, $\psi < 20\degr$. }
\label{plot:contact_line_heights}
\end{figure} 

It is desirable to express the maximal contact line height deformation $H$ in terms of the tilt angle $\psi$, since this is the parameter imposed in our simulations and would be easier to measure experimentally. Fig.~\ref{plot:contact_line_heights} shows the maximum contact line height difference, $H$, as a function of tilt angle, $\psi$, for a single isolated particle. $H$ is the difference between the maximum and minimum contact line heights, as shown in Fig.~\ref{plot:contact_line_heights}. We normalise the contact line height by $R_{\perp} (\alpha - 1)$, leading to a data collapse and corresponding scaling law for the small tilt-angle regime, $\psi < 20^{\circ}$. Therefore, in this regime we can write: 

\begin{align}
\label{eq:small_tilts_height}
H \simeq R_{\perp} (\alpha - 1)\psi.
\end{align}

The constant $-1$ in Eq.~\eqref{eq:small_tilts_height} takes into account the anisotropy of the particles: if the particles become spherical ($\alpha \to 1$) we expect the interface to remain completely flat as the particle tilts. Using this scaling law in Eq.~\eqref{eq:small_tilts_height} we can derive a pair potential for ellipsoidal particles with small tilt angles by substituting this expression into Eq. \ref{eqn:polardipolepotential_simple}. If we also define an average contact line radius $R_c = \frac{1}{2} (R_{\perp} + R_{\|})$ that takes into account the anisotropy of the ellipsoidal particle, we obtain:

\begin{equation}
E \simeq \frac{\pi}{4} \gamma_{12} R_{\perp}^4 (\alpha^2 - 1)^2   \frac{\psi^2}{r^2} 
\label{eqn:small_tilts}
\end{equation} 

\begin{figure}
 \includegraphics[width=\linewidth]{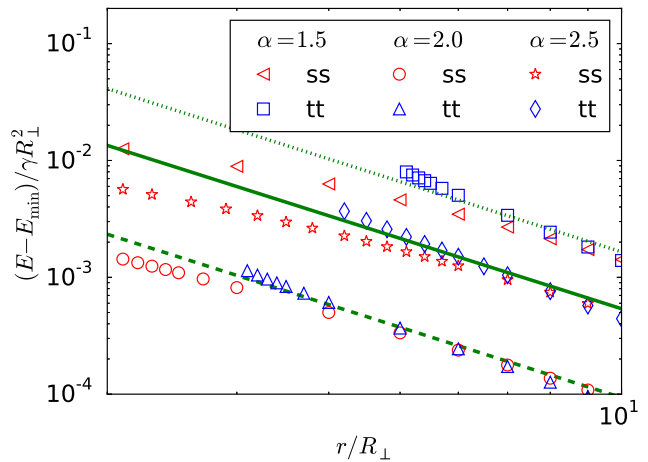}
 \caption{Dependence of the pair interaction energy on the particle centre-centre separation, $r$, for particles in the side--side (red symbols) and tip--tip orientation (blue symbols) for aspect ratios $\alpha = 1.5$, $2.0$, and $2.5$. The green lines are the predictions of our theoretical model in Eq.~\eqref{eqn:small_tilts}. We see good agreement between the model (lines) and the data (symbols) for large inter-particle separations, but strong deviations in the near-field due to higher order multipoles. The side--side state ($\psi=90$) is lowest in energy for all inter-particle separations.}
\label{plot:scaling}
\end{figure}

In Fig.~\ref{plot:scaling} we plot the capillary interaction energy between two ellipsoidal particles with tilt angle $\psi =5\degr$ as their centre-centre separation $r$ varies for different aspect ratios in both their side--side state (red symbols) and tip--tip state (blue symbols). For each configuration, we compare with our theoretical model in Eq.~\eqref{eqn:small_tilts} (green lines). 

We find good quantitative agreement between our theoretical model (Eq.~\eqref{eqn:small_tilts}) and our numerical data. Asymptotically, the interaction energy conforms to the $1/r^2$ power law predicted by Eq.~\eqref{eqn:small_tilts}.  We found similar agreement for tilt angles up to $\psi = 20\degr$, in accordance with the valid range of tilt angles for our scaling law in Eq.~\eqref{eq:small_tilts_height}. 

In the near-field (small inter-particle separations $r$), there is a strong deviation from the $1/r^2$ power law which we attribute to the importance of higher order multipoles that has also been observed for capillary quadrupoles.~\cite{lehle_ellipsoidal_2008, botto_capillary_2012} We find that the side--side configuration has a lower energy than the tip--tip configuration for all centre-centre separations, which corroborates our findings in Fig.~\ref{plot:representative_e} showing the side--side state to be the global energy minimum for particles in contact. 

The attractive force on each ellipsoidal particle is $-\frac{\partial E}{\partial r}$. From Fig.~\ref{plot:scaling}, we see that the attractive force at contact for the side--side orientation is smaller than that predicted by the superposition approximation, but of the same order of magnitude.  The superposition approximation can therefore give estimates of attractive forces for ellipsoidal particles inducing capillary dipoles with an $O(1)$ error, acceptable in many practical calculations. 

%%%%%%%%%%%%%%%%%%
%%%%%%%%%%%%%%%%%%

\begin{figure*}[t]
\centering
\subfloat[$N=3$]{
\begin{minipage}[b]{0.187\linewidth}
\includegraphics[width=\linewidth]{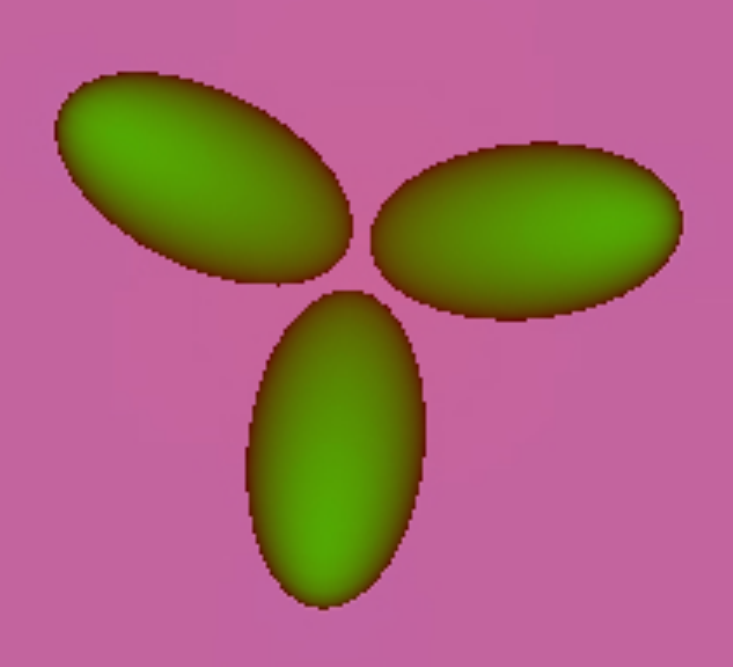} 
%\caption{$N=3$}
\label{fig:3p}
\end{minipage}
}
\subfloat[$N=4$]{
\begin{minipage}[b]{0.17\linewidth}
\includegraphics[width=\linewidth]{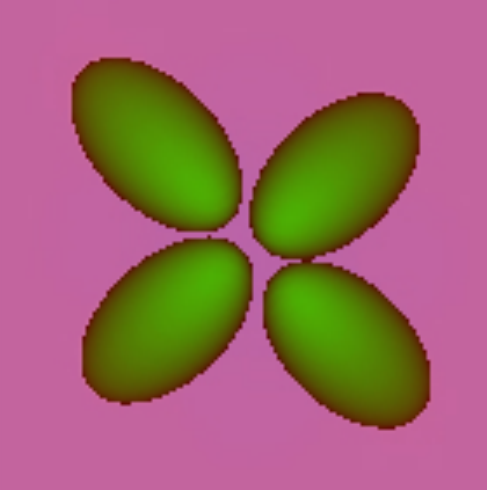} 
%\caption{$N=4$}
\label{fig:4p}
\end{minipage}
}
\subfloat[$N=5$]{
\begin{minipage}[b]{0.17\linewidth}
 \includegraphics[width=\linewidth]{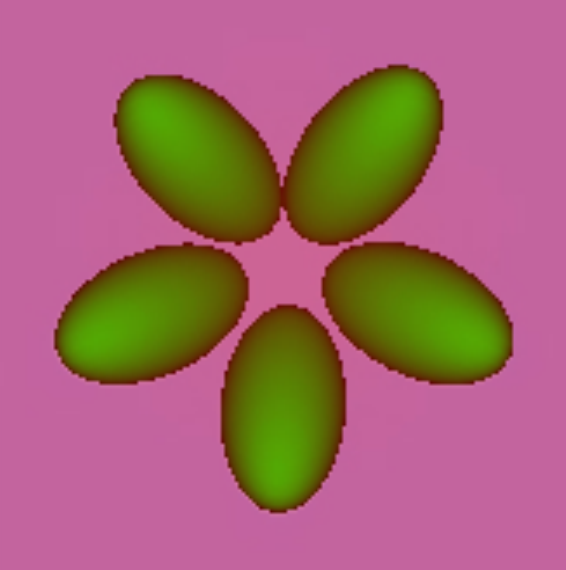} 
%\caption{$N=5$}
\label{fig:5p}
\end{minipage}
}
\subfloat[$N=6$]{
\begin{minipage}[b]{0.19\linewidth}
\includegraphics[width=\linewidth]{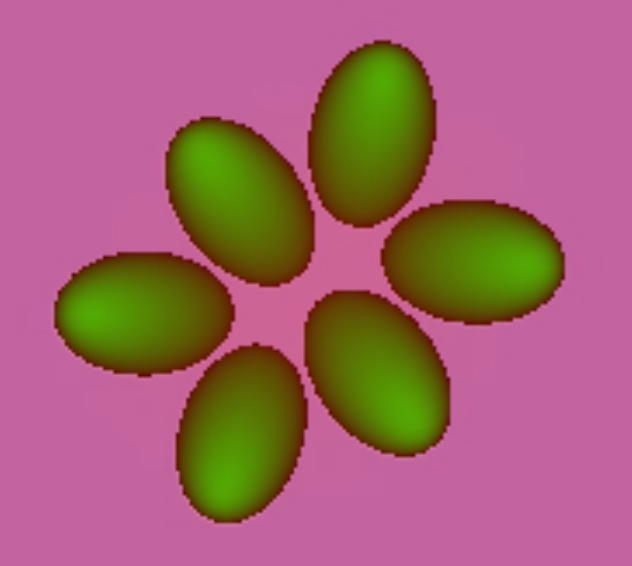} 
%\caption{$N=6$}
\label{fig:6p}
\end{minipage}
}
\subfloat[$N=12$]{
\begin{minipage}[b]{0.175\linewidth} 
\includegraphics[width=\linewidth]{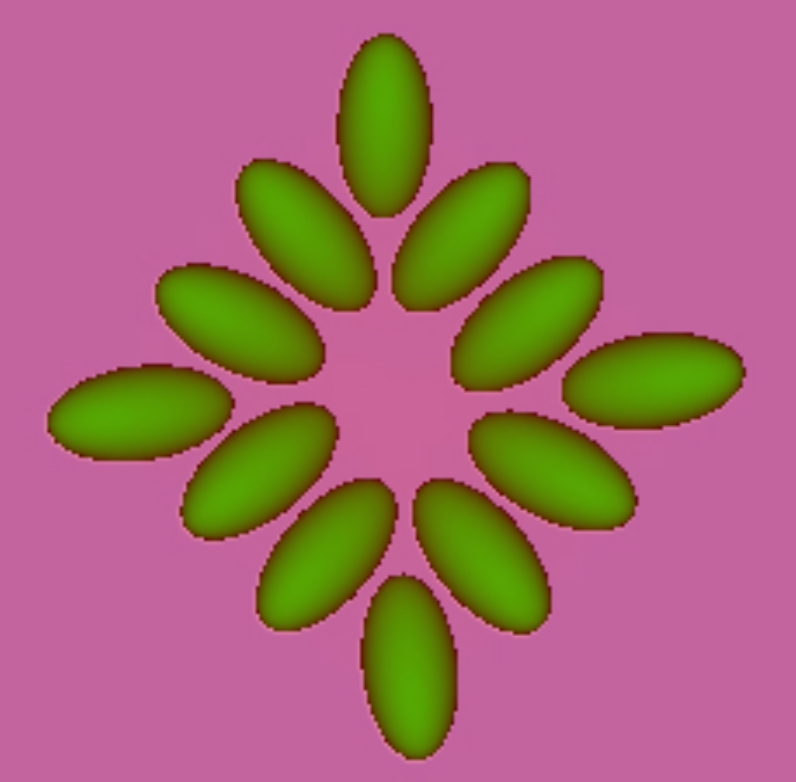} 
%\caption{$N=12$}
\label{fig:12p}
\end{minipage}
}
\caption{Lattice Boltzmann simulations: equilibrium clusters observed for aspect ratio $\alpha = 2.0$ and intermediate external field strength ($B=0.5B_c$). These novel clusters could form the basis for unique colloidal molecules. The fact that the particles do not form side--side chains shows that many-body interactions and the presence of higher-order multipoles for particles in contact significantly affect the interaction between tilted ellipsoidal particles.}
\label{fig:clusters}
\end{figure*}

Experimentally, capillary dipoles could be studied by using an external magnetic field to apply a torque to rod-like interfacial colloids. The strength of the field should be chosen to enforce a sufficiently large interface deformation for a given magnetic permittivity of the particles. Calling $\mathbf{M}_{e}$ the external torque on the particles, the condition for a sufficiently large deformation is obtained by equating the magnitude of the external torque ${M}_{e}$ to the capillary torque $\gamma_{12} L^2$ resisting tilting, where $L$ is the characteristic particle size. Taking $L=1\mu m$, and using the surface tension of a water-air interface we obtain ${M}_{e} \sim 7 \times 10^{-14} \mathrm{Nm}$. This torque should be achievable with a strong magnet (e.g. $B\simeq 100G$ and a magnetic moment of $3 \times 10^{-10}\mathrm{emu}$~\cite{lee_interfacial_2009}).    

In Eq.~\eqref{eqn:polardipolepotential} the capillary potential was calculated as a function of the dipole amplitude $H$. It is useful to express the interaction energy $E$ in terms of ${M}_{e}$ directly. To do so, we use the fact that  
$
E \simeq - \mathbf{P}_{A} \cdot \nabla h_B,
$
where $\mathbf{P}_{A}= \mathbf{e}_z \times \mathbf{M}_{e,A}$ is the capillary dipole moment induced by particle $A$, and $h_B$ is the interface deformation induced by $B$ evaluated at the centre of $A$;  $\mathbf{e}_z$ is the unit vector normal to the undisturbed interface.~\cite{hidalgo-alvarez_structure_2009} The field $h_B$ depends  linearly on the corresponding dipole moment according to   
$
 h_B = \frac{1}{2 \pi \gamma_{12} r} \mathbf{P}_B \cdot \hat{\mathbf{r}},
$
where $\hat{\mathbf{r}}$ is the unit vector along the line connecting $A$ to $B$.  The pair interaction is thus of the order of $E \sim \frac{1}{2\pi \gamma_{12} r^2}{M}_{e}^2$. For a particle of nominal radius $L$ the capillary interaction energy between two particles at contact  can thus be roughly estimated as 
$
E \sim \frac{1}{8\pi \gamma_{12} L^2}{M}_{e}^2.
$
We have $E \sim \gamma_{12} L^2$ when  ${M}_{e} \sim \gamma_{12} L^2$.  The predicted interaction energies can thus amount to  several millions $k_B T$ for micron-size particles and be substantial also for nanoparticles (for common fluid combinations $\gamma_{12}$ is in excess of $10 k_BT / \mathrm{nm}^2$). By tuning the ratio ${M}_{e}/(\gamma_{12} L^2)$, which is the characteristic Bond number for the torque, the dipolar capillary interaction energy can be reduced to any desired magnitude. 
\subsection{Many-Body Effects }
\label{sectionD}

It is interesting to compare the results and predictions of the current paper with the results of~\citet{davies_assembling_2014}, who studied the steady-state structures of monolayers of tilted ellipsoidal particles. They found that the particles had a tendency to form chains of particles in a side--side configuration, suggesting a deep energy minimum corresponding to that state. This result agrees with our simulation for pairs of ellipsoidal particles in contact (Fig.~\ref{plot:representative_e}). 

The curvature of the chains depended on the dipole-field strength and therefore the tilt angle (which are linearly related for sub-critical dipole-field strengths $B<B_c$~\cite{davies_interface_2014, bresme_orientational_2007}): the chains became straighter with increasing external field magnitude. This observation is compatible with our numerical results for particle pairs: the flexibility of a chain can be related to the curvature of the energy well for the side--side configuration.~\cite{botto_capillary_2012-1} Our results show that the energy well curvature increases as the tilt angle increases (Fig.~\ref{plot:representatives}). 

In addition to chains,~\citet{davies_assembling_2014} observed other local structures: particle triplets located at the intersection between two linear chains arranged with their tips close to each other so as to form  sharp ``bends''; particle triplets arranged in star-like structures; and closed rings formed by clusters of 7-8 particles. 

To study these structures without the complications associated with large particle numbers, we carried out lattice Boltzmann simulations of small clusters formed by $N=3$, $4$, $5$, $6$, and $12$ particles.  The particles have aspect ratio $\alpha=2.0$ and we applied an external tilting torque of magnitude half that required to make the particles flip into a vertical orientation with their major axes normal to the interface. Starting from a random initial placement of ellipsoidal particles adsorbed at the interface, we let the system achieve steady-state for several random initial configurations. We show the most frequently achieved steady-state structures in Figure~\ref{fig:clusters} for a given number of particles.  

For $N=3$, we find a star-like configuration characterised by ellipsoidal particles with their tips in contact and their axes diverging from a point. This structure, which evidently respects the symmetries of the 3-particle system, is similar in appearance to some of the local configurations seen in~\citet{davies_assembling_2014} Note that the ground state structure for three ellipsoidal particles interacting as capillary quadrupoles is a straight chain, closely followed by a triangular structure in which each particle tip is in contact with another particle tip.~\cite{loudet_self-assembled_2009} This configuration is different from that in Fig.~\ref{fig:3p}.

As the particle number increases to $N=5$, $6$, and $12$, we observe the formation of polygonal rings of particles.  For $N=12$, we observe a 4-sided structure: two ellipsoidal particles arranged in a side-by-side configuration form each side with a small bond angle between them. The ellipsoidal particles located at each of the four corners are slightly offset from the structure, forming a configuration that is reminiscent of the sharp ``bends'' found between rectilinear chains in Fig. 3 of~\citet{davies_assembling_2014}  Interestingly, we did not observe completely straight chains of side--side ellipsoidal particles for small particle numbers, even when simulated the particles initially arrange side--side rather than randomly. A larger number of particles is evidently needed for the closed chains observed in Fig.~\ref{fig:clusters} to open, producing the initial stage of formation of a percolating network. 

\section{Conclusions}

In this paper we numerically studied the interactions between ellipsoidal particles that tilt with respect to a fluid-fluid interface. Tilting induces dipolar interface deformations and corresponding dipolar capillary interactions. We showed that dipolar capillary interactions between pairs of ellipsoidal particles have unique features in comparison with the more studied quadrupolar capillary interaction (between ellipsoidal particles). 

For ellipsoidal particles, the side--side configuration is the global energy minimum for both quadrupolar and dipolar interactions, but we found that the dipolar interactions present an energy barrier that is absent in the case of quadrupolar capillary interactions.~\cite{botto_capillary_2012-1} Additionally, we found that the magnitude of this energy barrier increases, and the depth of the energy well for the metastable tip--tip configuration decreases, as the tilt angle increases. 

We developed a theoretical model describing the far-field interaction between two tilted ellipsoidal particles. We found excellent agreement between our model and the numerical data for large particle separations, validating our model, and strong deviations in the near-field as observed in previous studies of quadrupolar capillary interactions.~\cite{lehle_ellipsoidal_2008} 

Pair interaction results may be insufficient to describe the structures formed at fluid-fluid interfaces due to capillarity, because of the non-linear and many-body nature of these interactions. To get insights into many-body effects, we carried out lattice Boltzmann simulations of small clusters formed by  $3$, $4$, $5$, $6$, and $12$ ellipsoidal particles. For small clusters, the simulated arrangements have regular symmetries, and the structures we observe are similar to the local particle arrangements found in particle monolayers by~\citet{davies_assembling_2014} As the number of particles increases, polygonal rings appear to form. Therefore, as the surface coverage of particles increases, we expect a transition from a microstructure comprising isolated symmetric clusters to one comprising chains of different degrees of curvature. The high regularity of the clusters we obtain in simulations suggest that dipolar capillary interactions can potentially be used as building blocks to create planar ``colloidal molecules''.~\cite{manoharan_dense_2003, isa_particle_2010}

LB acknowledges EU funding from Marie Curie CIG grant FLOWMAT (618335).

\bibliography{bib}

\end{document}